\documentclass{article}
\usepackage{epsfig,multicol}
\usepackage{fleqn,espcrc2}
\title{\bf Under-doped La$_{2-x}$Sr$_x$CuO$_4$ with x = 0.063 - 0.125:\\ TSFZ growth of high-quality crystals and anomalous doping dependences of superconducting properties}

\author{F Zhou$^1$, W X Ti$^1$, J W Xiong$^1$ and Z X Zhao$^1$\\
X L Dong$^{1, 2}$, P H Hor$^2$, Z H Zhang$^2$ and W K Chu$^2$ \\
$^{1}$ National Laboratory for Superconductivity, Institute of Physics \& Center for Condensed Matter Physics, Chinese Academy of Sciences, P.O. Box 603, Beijing 100080, P. R. China\\
$^{2}$ Texas Center for Superconductivity and Department of Physics\\
University of Houston, Houston, Texas 77204-5002\\
\vspace{12pt}
\it to be published on Superconductor Science and Technology}

\begin{document}

\begin{abstract}
A series of high-quality La$_{2-x}$Sr$_x$CuO$_4$ (LSCO) superconductor crystals in the under-doping region with x = 0.063, 0.07, 0.09, 0.10, 0.111 and 0.125 has been successfully prepared by traveling-solvent floating-zone (TSFZ) technique. The crystals are large and free of sub-grains and foreign phases. The high crystal quality has been revealed by double-crystal x-ray rocking curves and Rutherford backscattering spectrometry combined with ion-beam channeling effect. We find that the evolutions of the superconducting transition width and volume fraction as a function of carrier concentration exhibit interesting anomalies in the vicinity of some "magic number" doping levels such as x = 1/4$^2$ (=0.0625) and 1/3$^2$ (=0.111). We argue that these behaviours are of intrinsic electronic origin.
\vspace{12pt}
\end{abstract}

\maketitle

La$_{2-x}$Sr$_x$CuO$_4$ (LSCO) system is known to be one of the cuprate superconductor systems with fewer components and a simple layered structure of K$_2$NiF$_4$ type which can be basically described as a stacking of single-plane CuO$_2$ superconducting layers separated by two (La, Sr)O layers. The intrinsic Josephson junctions of LSCO have recently been characterized as stacked serial SIS junctions \cite{1}, and its high Josephson plasma frequency in THz region is quite interesting for developing new electronic devices such as THz generators and detectors \cite{2}. On the other hand, its electronic states strongly depend on hole-doping which can be achieved with the substitution of Sr$^{2+}$ cations for La$^{3+}$ ones. At very low doping (x $<$ 0.02), La$_{2-x}$Sr$_x$CuO$_4$ is an antiferromagnetic Mott insulator. After crossing a spin glass state (0.02 $<$ x $<$ 0.05) \cite{3,4}, it becomes a superconductor (0.05 $<$ x $<$ 0.26) and then changes to a normal metal (0.26 $<$ x $<$ 0.6) \cite{5} and finally it behaves as a semiconductor (x $>$ 0.6) \cite{6}. In the superconductor regime, the critical temperature of LSCO reaches its maximum of 38 K at x = 0.15. The investigation on the doping dependences of low temperature physical properties in both normal and superconducting states is important for acquiring a better understanding of high T$_C$ superconductivity and strongly correlated electron systems. LSCO system is therefore not only attractive in its possible applications but also in such fundamental research. For both purposes, high-quality and sizable single crystals are highly demanded, and persistent effort has been made in the crystal growth using different techniques \cite{7,8,9,10,11,12}. Traveling-solvent floating-zone (TSFZ) method is widely accepted as a unique approach to growing such incongruent melting oxide superconductor crystals. This crucible-free technique avoids the crystal contamination from crucible material that will otherwise interfere with the observation of intrinsic physical properties. The large temperature gradient produced by focused lamp images makes possible good solid-liquid interfaces and a stable molten zone.

By using TSFZ method, we have successfully prepared a series of large and high-quality La$_{2-x}$Sr$_x$CuO$_4$ single crystals covering some interesting under-doping levels (x = 0.063 - 0.125) with the aim of systematic investigation on the physical properties. We first report here the results of TSFZ growth and characterizations of some properties of these LSCO crystals. Interesting anomalous doping dependences of superconducting properties are observed in this series of crystals and are briefly discussed.

High purity ($\geq$ 99.99\%) oxides La$_2$O$_3$, CuO and carbonate SrCO$_3$ were used as the raw materials. For the feed rods, the initial composition was (1-x/2)La$_2$O$_3$/xSrCO$_3$/CuO in molar ratio with x = 0.063, 0.07, 0.09, 0.10, 0.111 and 0.125. Excess CuO of 1 - 2 mol\% was added for compensating its evaporation loss in the growth process. The solvents were much richer in CuO as self-flux, typically of 78 mol\% CuO. In determining the Sr contents of the solvents, we took into account the previously reported results of distribution coefficients {\it k$_{Sr}$} of Sr doping into La$_2$CuO$_4$ \cite{10}. The Preparation of dense and homogeneous feed rods is one of the key factors in achieving a stable and continuous TSFZ growth. Each step in the preparation was carefully checked and optimized, such as thorough mixing and grinding of the starting and prefired powders by a planetary ball mill, forming compact cylindrical rods under a high hydrostatic pressure before the feed rods being finally well sintered. An infrared heating floating-zone furnace with a quartet ellipsoidal mirror (Crystal Systems Inc., Model FZ-T-10000-H) was used for TSFZ experiments. The crystals were grown under an oxygen pressure of 0.2 MPa at a zone traveling rate of 0.8 mm/hr using seed crystals orientated along orthorhombic or tetragonal [100] or [110] directions.

By careful operations and the use of high quality feed rods, a stable molten zone was successfully maintained in most experiments until proceeding growths were artificially stopped just before the whole feed rods, typically 125 mm long, ran out. Shown in Fig.1 is the photo for one of as-grown ingots, with a typical size of 5-6 mm in diameter and 110 mm in length. Polished crystal pieces cut from the ingots were checked by optical microscopy using polarized or normal light. No grain boundaries were observed in the crystallized ingots after several centimeters away from the starting point of crystallization on seed crystals, indicating that large single-grain crystals were obtained. The compositions of grown crystals estimated by ICP-AES were very close to those of feed rods, and no foreign phases in crystal samples were detected by powder XRD analysis.

\begin{figure}[tbh]  
\begin{center}
\epsfig{file=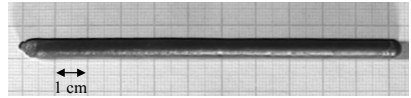,bb=176 579 418 639,width=5.4cm,clip} 
\caption{As-grown ingot of La$_{1.875}$Sr$_{0.125}$CuO$_4$ crystal.}
\end{center}
\end{figure}

Experiments of x-ray rocking curves were performed on representative La$_{1.91}$Sr$_{0.09}$CuO$_4$ (x = 0.09) crystal using a double-crystal diffractometer (Rigaku, Model SLX-1) equipped with a 12 KW rotating Cu target and a Ge (004) monochromator ($\alpha \sim 5^o$). Fig. 2 shows the rocking curve of (008) Bragg reflection for this sample having a surface dimension of 4mm $\times$ 15mm (the beam slit size was set to 0.5mm $\times$ 10mm). The full-width-at-half-maximum (FWHM), which is correlated with the crystal mosaicity, is as small as 0.10$^o$. This is one of the best data reported so far for LSCO crystals \cite{12}. The quality of the same crystal was also examined by Rutherford backscattering spectrometry combined with ion-beam channeling effect (RBS-channeling) which was carried out using 2 MeV well collimated He ion beam and a Si detector set at 160$^o$ to the incident direction. The channeling spectrum was taken along the c-axis of the crystal. The RBS-channeling minimum yield \cite{13} is only $\chi_{min}$ = 3.8\% (Fig. 3), being another evidence for the high quality of the crystal. Moreover, it can be seen from Fig. 3 that, for the aligned spectrum, the backscattering counts increase very slowly with depth. This is a strong indication that the defect density in the crystal is very low.

\begin{figure}[tbh]  
\begin{center}
\epsfig{file=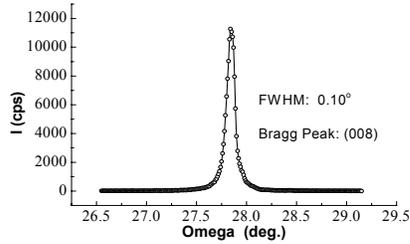,bb=176 504 420 651,width=5.4cm,clip} 
\caption{Rocking curve of (008) reflection for La$_{1.91}$Sr$_{0.09}$CuO$_4$ (x = 0.09) crystal
taken by double-crystal x-ray diffraction using Cu K$\alpha_1$ radiation. The surface dimension of the sample is 4mm $\times$ 15mm and the beam slit size is 0.5mm $\times$ 10mm.}
\end{center}
\end{figure}

\begin{figure}[tbh]  
\begin{center}
\epsfig{file=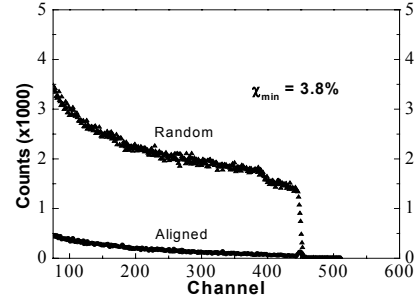,bb=175 472 421 654,width=5.4cm,clip} 
\caption{2-MeV $^4$He$^+$ ions RBS-channeling effect on La$_{1.91}$Sr$_{0.09}$CuO$_4$ (x = 0.09)
crystal. The aligned spectrum is taken along c-axis of the crystal and the random spectrum with the sample rotating and tilted 10 degrees away from the aligned direction. The RBS-channeling minimum yield is only $\chi_{min}$ = 3.8\%.}
\end{center}
\end{figure}

The Meissner (field-cooled) and shielding (zero-field-cooled) signals of all the six crystal samples were measured on a SQUID magnetometer (Quantum Design, MPMS XL) in a low field of 5 Oe with c//H, as shown in Fig. 4. The superconducting critical temperatures observed from the Meissner effects range from about 11K for x = 0.063 to about 31K for x = 0.125. It is of much interest to note that, around some special doping levels with "magic numbers" such as x = 1/4$^2$ (=0.0625) and 1/3$^2$ (=0.111), the superconducting transitions are much sharper ($\Delta$T $\approx$ 2K) than those at the doping contents away from them and the Meissner fraction drops remarkably upon doping in the close vicinity of x = 1/3$^2$ (Fig. 4). Moreover, the shielding signals exhibit the same broadening trend for the doping levels away from the magic numbers and they all have 100\% volume fraction. This indicates that there is no macroscopic inhomogeneity and/or weak links. Generally speaking, crystalline imperfection may broaden superconducting transition and a narrow transition width is expected for superconductor crystals of high quality. We have evidenced that the x = 0.09 LSCO crystal is of excellent crystalline perfection, but, contrary to the expectation, it exhibits a broad superconducting transition. As a further check, the La$_{1.91}$Sr$_{0.09}$CuO$_4$ sample was annealed under various conditions for several to twelve days at 600-1000$^o$C under oxygen flow or 0.3 MPa oxygen pressure before quenched in liquid nitrogen. However, magnetic measurements showed that its transition width remained unchanged after these annealing processes. These results and observations thus lead us to conclude that such anomalies of superconducting properties in the vicinity of these magic doping levels as well as the broad transitions at the dopant contents away from these magic numbers are of intrinsic electronic origin. At these special carrier concentrations, some specific electronic state may occur but it may not be completely pinned down elsewhere in this doping region. Indeed, both superconducting electronic phase separations \cite{14} and the formation of 2D electronic lattices \cite{15,16,17} at magic doping numbers  are reported in Sr doped and Sr/O co-doped lanthanum cuprates. Further related investigations are currently in progress.

\begin{figure}[tbh]  
\begin{center}
\epsfig{file=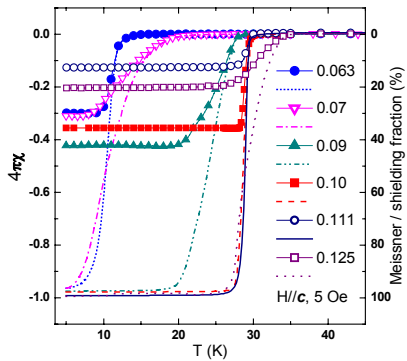,bb=180 456 414 664,width=5.4cm,clip}
\caption{Meissner curves (symboled lines) and shielding curves (bare lines) of La$_{2-x}$Sr$_x$CuO$_4$ single crystals with various doping levels (x = 0.063 - 0.125).}
\end{center}
\end{figure}

In conclusion, we have successfully prepared by TSFZ method a series of high-quality La$_{2-x}$Sr$_x$CuO$_4$ superconductor crystals in the under-doping regime with x = 0.063 - 0.125. The crystals are pure, subgrain-free and large in size. We have observed from the magnetic measurements that the superconducting transition width narrows in the close vicinity of some magic doping levels while remains broad at other dopant contents, and the Meissner fraction shows an unusual decrease around x = 1/3$^2$. We suppose that such interesting anomalous behaviours in superconducting state may probably originate from some intrinsic electronic state occurring at these special carrier-dopings.

{\bf Acknowledgement} The authors at IP\&CCMP-CAS are very
grateful to Prof. Z. L. Wang and Mrs. N. L. Li for their various
and valuable helps in crystal orientation and processing. They
would also like to thank many other colleagues for their
technical assistances and measurements (SQUID, XRD, SEM etc.).
The work in Beijing is supported by {\it Ministry of Science and
Technology of China and National Natural Science Foundation of
China} through Project G1999064601 and Project 10174090. The work
in Houston is supported by {\it the State of Texas through The
Texas Center for Superconductivity}.


\begin{thebibliography}{50}
\bibitem{1} Uematsu Y, Mizugaki Y, Nakajima K, Yamashita T, Watauchi S and Tanaka I 2002 Physica C 367 382).
\bibitem{2} Yamashita T 2001 Physica C 362 58.
\bibitem{3} Wakimoto S, Shirane G, Endoh Y, Hirota K, Ueki S, Yamada K, Birgeneau R J, Kastner M A, Lee Y S, Gehring P M and Lee S H 1999 Phys. Rev. B 60 R769.
\bibitem{4} Chou F C, Belk N R, Kastner M A, Birgeneau R J, Aharony A 1995 Phys. Rev. Lett. 75 2204.
\bibitem{5} Torrance J B, Yokura Y, Nazzal A I, Bezinge A, Huang T C and Parkin S S P 1988 Phys. Rev. Lett. 61 1127.
\bibitem{6} Sreedhar K and Ganguly P 1990 Phys. Rev. B 41 371.
\bibitem{7} Tanaka I and Kojima H 1989 Nature 337 21.
\bibitem{8} Chen C, Watts B E, Wanklyn B M, Thomas P A and Haycock P W 1988 J. Crystal Growth 91 659.
\bibitem{9} Cassanho A, Keimer B and Greven M 1993 J. Crystal Growth 128 813.
\bibitem{10} Kojima H, Yamamoto J, Mori Y, Khan M K R, Tanabe H and Tanaka I 1997 Physica C 293 14.
\bibitem{11} Marin C, Charvolin T, Braithwaite D and Calemczuk R 1999 Physica C 320 197.
\bibitem{12} Komiya S, Ando Y, Sun X F and Lavrov A N 2002 Phys. Rev. B 65 214535.
\bibitem{13} Lindhard J 1965 Mat. Fys. Medd. K. Dan. Vidensk. Selsk. 34.
\bibitem{14} Lorenz B, Li Z G, Honma T and Hor P H 2002 Phys. Rev. B 65 144522.
\bibitem{15} Kim Y H and Hor P H 2001 Modern Physics Letter B 15 497.
\bibitem{16} Hor P H and Kim Y H 2002 J. Phys.: Condens. Matter 14 10377.
\bibitem{17} Kim Y H, Hor P H, Dong X L, Zhou F, Zhao Z X, Song Y S, to be submitted.
\end{thebibliography}
\end{document}